\documentclass[sigconf]{acmart}
\pdfoutput=1
\usepackage{booktabs}
\usepackage{color}
\usepackage{comment}
\usepackage[utf8]{inputenc}
\usepackage{fullpage} 
\usepackage{graphicx}
\usepackage{hyperref} 
\usepackage{listings}
\usepackage{subfigure}

\hypersetup{colorlinks=true,citecolor=[rgb]{0,0.6,0}}
\definecolor{whitegray}{rgb}{0.985,0.985,0.975}

\setcopyright{iw3c2w3}
\acmConference[WWW'18 Companion]{The 2018 Web Conference Companion}{April 23--27, 2018}{Lyon, France}
\acmYear{2018}
\acmPrice{}
\acmDOI{10.1145/XXXXXX.XXXXXX}
\acmISBN{978-1-4503-5640-4/18/04.}

\begin{document}

\title{Linking ImageNet WordNet Synsets with Wikidata}
\author{Finn Årup Nielsen}
\affiliation{%
  \institution{Technical University of
    Denmark}
  \department{DTU Compute}
  \streetaddress{Richard Petersens Plads}
  \state{}
  \postcode{2800}
  \city{Kongens Lyngby}
  \country{Denmark}
}
\email{faan@dtu.dk}

\thanks{I would like to thank Laura Rieger and Lars Kai Hansen for
  discussions. This research was supported by Innovation Fund Denmark
  through the DABAI project.}

\begin{abstract}
  The linkage of ImageNet WordNet synsets to Wikidata items will
  leverage deep learning algorithm with access to a rich multilingual
  knowledge graph.
  Here I will describe our on-going efforts in linking the two
  resources and issues faced in matching the Wikidata and WordNet
  knowledge graphs. 
  I show an example on how the linkage can be used in a
  deep learning setting with real-time image classification and
  labeling in a non-English language and discuss what opportunities
  lies ahead.
\end{abstract}


\maketitle

\section{Introduction}

Machine learning with deep neural networks has in recent years yielded
interesting advances with a range of tasks.
Parti\-cularly image classification models have seen advances with
the first system claiming ``superhuman visual pattern recognition'' in
a restricted and controlled setting back in 2011 \cite{Q34451415}.
More recent trends have explored how knowledge bases can be utilized
together with machine learning.
For instance, a combined word embedding of the ConceptNet knowledge graph
and the two popular word embeddings, word2vec and GloVe, could
produce state-of-the-art results on word similarity tests
\cite{Q31895639},
while the Wembedder system uses an embedding of the Wikidata knowledge
graph in a web application with recommender system features \cite{Q41799598}.

To take advantage of knowledge graphs in machine learning we should
deep-link systems, and preferably not just on the word-level but on a
semantic level where the different senses of polysemic words are
distinguished. 
One opportunity is a linkage between ImageNet through Wordnet to
Wikidata. 
Below I will describe these three resources and follow with a detailed
account of issues around on-going work on matching the items across
these resources. 
I will show statistics and describe a small machine learning
application that uses the linkage between ImageNet and Wikidata, and
lastly discuss issues concerning the matching and the opportunities for
extending Wikidata.

\section{WordNet, ImageNet and Wikidata} 
\subsection{WordNet}

WordNet is a machine readable lexical resource that describe words.
Words that are synonyms are grouped into items called
\emph{synsets} \cite{Q36655252}.
Each WordNet synset is associated with an Semantic Web Linked Open
Data (LOD) URI.
URIs for WordNet for version 3.0 has been prefixed with
\url{http://wordnet-rdf.princeton.edu/wn30/} while URIs for version
3.1 synsets are prefixed with \url{http://wordnet-rdf.princeton.edu/wn31/}.
The canonical prefix has been change recently, so the part
\verb!wn30! has been changed to \verb!pwn30!.\footnote{See the
  discussion on the WN-USER mailing list at
  \url{https://lists.princeton.edu/cgi-bin/wa?A2=ind1801&L=wn-users&F=&S=&P=1719}. The
old URIs are redirected.}

\subsection{ImageNet}

ImageNet \cite{Q37043902} is a large collection of 
images, labeled against WordNet 3.0 and described at
\url{http://image-net.org/}.
According to January statistics, ImageNet contains 
14,197,122 images and  21,841 indexed synsets.
There are various labeling schemes.
The object attributes scheme labels 400 synsets across 25 attributes.
The attributes are with respect to color, pattern, shape and
texture.\footnote{\url{http://image-net.org/download-attributes}} 
ImageNet Large Scale Visual Recognition Challenge ILSVRC \cite{Q47451393} is
an image recognition challenge that started in 2010 and has been
repeated for several years.
It has had several tasks: image classification, single-object
locatization or object detection.
For the image classification task a subset with 1,000 object categories
from ImageNet was selected corresponding to 1,000 WordNet synsets.
ILSVRC synsets have changed through the years so
only 639 have been used in all the years.

Inspired by the demonstration of the so-called AlexNet
\cite{Q30077834} where state-of-the-art image classification results
were reported on the ILSVRC dataset,
many publications describe the usage of the ILSVRC data for
training deep learning image classification models.
Researchers distribute pre-trained deep learning models based on the
ILSVRC data.
One such model is Resnet-50 described by \cite{Q35464016}.
High-level deep learning Python frameworks, Keras and Caffe, are able to
read this model in a few lines of code.
For instance, with Keras the pre-trained Resnet-50 model is set up
with \verb!model = ResNet50()!.\footnote{\url{https://keras.io/applications/\#resnet50}}
The Keras implementation will also convert the output of the
trained model from its index-based value to the WordNet synset
identifier. 

\sloppy
In ImageNet the synset identifier is represented with the word class
as a prefix, e.g., the synset identifier for banana is n07753592. 
So a (trivial) conversion is necessary to match the LOD URIs which in
the banana case would be 
\url{http://wordnet-rdf.princeton.edu/wn30/07753592-n}, while a
preview of corresponding ImageNet images is available as \break
\url{http://image-net.org/explore.php?wnid=n07753592}.

\subsection{Wikidata}

The collaborative knowledge graph, Wikidata, at
\url{https://www.wikidata.org} \cite{Q18507561}
has over 40 million items
(with the identifier prefixed with ``Q''),
that each may be described by several thousand properties (prefixed
with ``P'').
Wikidata properties with the data type ``External
identifier'' is often suggested as a means to deep linking external
databases. 
Such a method could also be applied for WordNet synset identifiers.
This requires a property suggestion and some deliberation among the
Wikidata editor before the external identifier property is made
available for use.\footnote{I suggested such a property, see
  \url{https://www.wikidata.org/wiki/Wikidata:Property_proposal/Wordnet_synset_ID}.}
There is another way to link Wikidata with WordNet.
Wikidata has a property to link to LOD URIs: the ``exact match''
property (P2888), corresponding to skos:exactMatch.
This method can be used to link a Wikidata item to its associated
WordNet synset, e.g.,
\emph{mashed potato} (Q322787) is linked via the P2888 property to
\url{http://wordnet-rdf.princeton.edu/wn30/07711569-n}.

Wikidata links to another lexical resource:
The multilingual BabelNet\footnote{\url{http://babelnet.org/}.} is
linked via its own external identifier property, the
P2581\footnote{\url{https://www.wikidata.org/wiki/Property:P2581}.}
which currently is used 59.105 times.\footnote{Wikidata Query Service can be
  queried with
  ``\texttt{SELECT (COUNT(*) AS ?count) WHERE \{ [] wdt:P2581 []
    \}}'', see \url{http://tinyurl.com/y8pk2cqa}.
}
While WordNet is one of the resources that uses BabelNet
\cite{Q47485515}, the identifier is different.
For instance, ``quill pen'' is
\href{http://babelnet.org/synset?word=bn:00065709n}{00065709n} in
BabelNet and 04033901-n in WordNet 3.0.

Once WordNet 3.0 URIs have been entered into Wikidata,
we can obtain the
inverse mapping from the WordNet URI to the Wikidata item
with the Wikidata Query Service (WDQS) SPARQL engine, e.g.,
the SPARQL query \break
``\texttt{SELECT * WHERE \{
?item wdt:P2888
\newline
<http://wordnet-rdf.princeton.edu/wn30/04033901-n>
\}}''
will identify
\href{https://www.wikidata.org/wiki/Q4063215}{Q4063215} as the Wikidata
item corresponding to the WordNet concept ``quill''. 

One advantage with Wikidata is its multilingual nature:
Once we have linked a WordNet synset to a Wikidata item we can record
labels in multiple languages, and in many cases the labels are already
present. For instance, ``quill pen'' is currently labeled for 33 languages.
The corresponding number for BabelNet is 12.
It is important to note though that the multilinguality in Wikidata is
certainly not complete and biased towards the major European languages
\cite{Q37859976}.

Wikidata and WordNet have been defined mostly independent of each
other. 
There are other linked semantic resources where this is also the case.
Examples are some of the wordnets inspired by the original English
(Princeton) WordNet \cite{Q37052366}.

\begin{table*}[t]
  \centering
  \def\arraystretch{1.5}
  \begin{tabular}{p{5.2cm}p{5.1cm}p{5.2cm}}
    \toprule
    ImageNet & WordNet & Wikidata  \\
    \midrule

    \href{http://image-net.org/explore.php?wnid=n07930864}{n07930864}:
    Mostly (coffee) cups, i.e., ``a small open container usually used
    for drinking; usually has a handle'', but also a few champagne
    glasses. 
    &
    \emph{Cup}; ``A punch served in a pitcher instead of a punch
    bowl'', --- the 6th sense of \emph{cup}.
    & ?
    \href{https://www.wikidata.org/wiki/Q1121224}{Q1121224}
    is an item for the parent synset (punch). 
    There seems to be none for the punch cup synset.
    \\

    \href{http://image-net.org/explore.php?wnid=n07565083}{n07565083}:
    Printed cards with information about dishes served.
    &
    \emph{Menu}: ``The dishes making up a meal''.
    & ?
    ImageNet-consistency could link to
    \href{https://www.wikidata.org/wiki/Q658274}{Q658274}.
    There does not seem to be an item
    corresponding to the WordNet sense.
    \\
    \href{http://image-net.org/explore.php?wnid=n03742115}{n03742115}:
    Image of bathroom cabinets.
    &
    \emph{Medicine chest}, medicine cabinet; ``Cabinet that holds
    medicines and toiletries'' 
    & Rather than medicine chest, the WordNet
    synset corresponds to bathroom cabinet
    (\href{https://www.wikidata.org/wiki/Q4869069}{Q4869069})
    \\
    \href{http://image-net.org/explore.php?wnid=n03832673}{n03832673}:
    Images linked
    directly to n03832673 show computer notebooks, but the \emph{planner} child     synset shows paper notebooks.    
    &
    \emph{Notebook} is a notebook computer, but the WordNet hierarchy
    displayed on the ImageNet homepage shows \emph{planner}
    \href{http://image-net.org/explore.php?wnid=n03956785}{n03956785}
    as a hyponym and this synset is described as ``A notebook
    for recording appointments and things to be done, etc.''
    There is a separate item for \emph{laptop}.
    & ? The closest Wikidata item may be
    \href{https://www.wikidata.org/wiki/Q3962}{Q3962} with
    \emph{laptop} as the current English label. 
    There does not seem to be an individual item for computer
    notebook. 
    \\
    \href{http://image-net.org/explore.php?wnid=n04152593}{n04152593}:
    Cathode-ray tube (CRT) screens as well as some flatscreens.
    Often the full apparatus, --- not just the display. Mostly
    computer screens.  
    &
    \emph{Screen}, CRT screen: ``The display that is
    electronically created on the surface of the large end
    of a cathode-ray tube''. 
    &
    ? \href{https://www.wikidata.org/wiki/Q5290}{Q5290} is
    computer monitor or screen, and not necessarily
    CRT-based. 
    \href{https://www.wikidata.org/wiki/Q1736293}{Q1736293}
    is cathode ray tube screen fitting WordNet but less so
    ImageNet. 
    \\
    \href{http://image-net.org/explore.php?wnid=n04355933}{n04355933}:
    Mostly (all?) protective eyewear.
    & \emph{Sunglass}; ``A convex lens that focuses the rays of the
    sun; used to start a fire''  
    & ?
    \href{https://www.wikidata.org/wiki/Q368027}{Q368027}
    is ``burning glass'' corresponding to the
    WordNet sense, but quite different from
    the examples in ImageNet.
    \\
    \href{http://image-net.org/explore.php?wnid=n03944341}{n03944341}:
    Toy pinwheels
    & \emph{Pinwheel}; ``A wheel that has numerous
    pins that are set at right angles to its rim''.
    There is another WordNet synset for the toy:
    \href{http://image-net.org/synset?wnid=n03944138}{n03944138}
    also called ``pinwheel wind collector''.
    & ? Pinwheel is a disambiguation page in
    English Wikipedia, while ``Pinwheel
    (toy)'' corresponds to ImageNet usage and
    n03944138.
    \\
    \bottomrule
  \end{tabular}
  \caption{Matching between ImageNet, WordNet and Wikidata. Some
    examples of the difficulties with aligning the resources.}
  \label{tab:matching}
\end{table*}

\section{Matching Wikidata with ImageNet}

For matching Wikidata with ImageNet, I focused on the ILSVRC-part of
ImageNet synsets
linking them up with the P2888 property and the
\verb!http://wordnet-rdf.princeton.edu/wn30/! prefix by examining
Wikidata definitions, WordNet synsets descriptions, ImageNet images
linked to the WordNet synsets and sometimes the description in
Wikipedia articles associated with Wikidata items. 
Matching Wikidata with ImageNet is for some concepts straightforward.
For instance, many taxons can readily be matched. 
At other times the matching requires some thought.
Table~\ref{tab:matching} shows examples where a match is
particular difficult.
Below I attempt a classification of various matching problems faced
during manual linking of Wikidata items to WordNet synsets.

An ILSVRC synset is {\bf missing from Wikidata}.
In most of these cases, it is easy to create new Wikidata items and link
them to WordNet.
Examples where I could not find matching items and created new items
are \emph{soup bowl} and \emph{bath towel}.

The matching Wikidata item is a Wikipedia {\bf disambiguation page}:
An example is car mirror where the English Wiki\-pedia page ``Car
mirror''  currently is a disambiguation page linking to two real articles
``Rear-view mirror'' and ``Wing mirror''.
A solution to such a problem would be to either create a new
Wikidata item for car mirror or change the definition of ``Car
mirror'' in both Wikidata and the English Wikipedia.
``Armchair is a disambiguation page in the English Wikipedia, but
another item link armchair in Wikidata
(\href{https://www.wikidata.org/wiki/Q25503439}{Q25503439}). 
\emph{Mixing bowl} is another example where the Wikidata item is
describing a disambiguation page.

There is a {\bf discrepancy between ImageNet and WordNet}:
For instance, 
\emph{sunglass} (\url{http://wordnet-rdf.princeton.edu/wn30/04355933-n})
is described as a ``a convex lens that focuses the rays of the sun;
used to start a fire''. However, the associated ImageNet images are
better described with the
\url{http://wordnet-rdf.princeton.edu/wn31/104296228-n} synset (dark
glasses, shades, sunglasses) described with ``spectacles that are
darkened or polarized to protect the eyes from the glare of the sun''.
This issue was previously noted by Tom White on Twitter in September
2017.\footnote{Tom White (@dribnet). aha: ``sunglass'' is ``burning
  glass'', \ldots
  \url{https://twitter.com/dribnet/status/904133638257655808} (3
  September 2017).}
In the same context he also reported differences between the 
\emph{sunglass} and the \emph{sunglasses} images.
A similar confusion occurs for \emph{maillot}, where Wordnet has two
synsets: ``a woman's one-piece bathing suit'' and ``tights for dancers
or gymnasts''.
For the latter synset, Image\-Net displays images for
one-piece bathing suits.
Another example is \emph{monitor} described as ``Electronic equipment
that is used to check the quality or content of electronic
transmissions''.
At ImageNet's \href{http://image-net.org/explore.php?wnid=n03782006}{n03782006},
the images shown are broader with many displaying computer monitors,
--- in Wikidata it would correspond to approximately the more general
concept \href{https://www.wikidata.org/wiki/Q6021804}{Q6021804}
(``electronic visual display'').
Related to the \emph{monitor} issue is \emph{screen} with the alias
\emph{CRT screen} and description ``The display that is electronically
created on the surface of the large end of a cathode-ray tube''. 
The associated ImageNet images show many screens that are not
CRT-based, but flat panel displays.
One interesting discrepancy is for \emph{spider web}:
WordNet has two synsets for that word: ``a web spun by spiders to trap
insect prey'' and ``a web resembling the webs spun by spiders'' where
the latter is regarded as an artifact and the former is not found in
ImageNet. 
ImageNet images labeled with the latter sense
(\href{http://image-net.org/explore.php?wnid=n04275548}{n04275548})
show real natural spider webs.
It is obvious that webs spun by spiders resemble webs spun by spiders,
but in this case the webs spun by spiders should probably not be
assigned to the WordNet synset.
Table~\ref{tab:matching} shows other examples of this problem,
e.g., for \emph{cup} (coffee cup or punch), \emph{menu} (printed list
or a set of 
servings), \emph{notebook} (computer or paper book for notes) and
\emph{pinwheel}. 
Not shown in the table is \emph{picture rail} (``rail fixed to a wall
for hanging pictures'') and its parent
\emph{rail} (``a horizontal bar (usually of wood or metal)''). 
For both these synsets ImageNet associates images of railways and trains.
These cases may involve concepts with a large semantic difference.
There are likely more discrepancies. 

There are some {\bf difference between the WordNet and the Wikidata
concepts with similar names}.
For instance, WordNet defines \emph{street sign} as ``a sign visible
from the street'' while Wikidata describes the same English word as a ``type
of traffic sign used to identify named roads''
(\href{https://www.wikidata.org/wiki/Q1969455}{Q1969455}).
The Wikidata item for traffic sign
(\href{https://www.wikidata.org/wiki/Q170285}{Q170285}) is described as
a ``symbol for people out in traffic'' corresponding better to the
\emph{street sign} of WordNet.
Another example is ``tea chest'' that in the English Wikipedia (the
only Wikipedia describing the concept) is explained as a case
``approximate size 500x500x750 millimeters'' with metal edges.
Wikidata links this concept to the identifier 300039143 of the \emph{Art \&
Architecture Thesaurus} where it is explained as ``Square wooden cases
usually fixed with sheet lead or tin and used for exporting tea.''
The English Wikipedia further extends the concept with ``The term is
now used more widely to indicate similarly-sized cases, including
cardboard cases, produced for various home and commercial uses.'' 
In ImageNet \emph{tea chest} (n04397168) are cases with tea of varying
sizes corresponding well to the WordNet's general description:
``Chest for storing or transporting tea''.
Here we have an example of the English Wikipedia describing two
concepts in the same article, a specific and a general, where the
general topic correspond to WordNet and ImageNet concepts.

There are a number of other cases where the matching requires an
effort to resolve.
These may involve  {\bf multiple semantically
  similar items in WordNet and Wikidata}:
For instance, there are multiple synsets for \emph{radiator}.
The \emph{radiator} synsets in ILSVRC
(\href{http://image-net.org/synset?wnid=n04040759}{n04040759}) is
described as ``A mechanism consisting of a metal honeycomb through
which hot fluids circulate; heat is transferred from the fluid through
the honeycomb to the airstream that is created either by the motion of
the vehicle or by a fan''.
Most of the ImageNet images associated with this synset fit the
description of another synset described as a ``heater consisting of a
series of pipes for circulating steam or hot water to heat rooms or
buildings''.
In Wikidata there are 3 or 4 items close in semantics to
\emph{radiator}.
WordNet distinguishes between \emph{tape player},
(\href{http://image-net.org/synset?wnid=n04392985}{n04392985}) and
\emph{tape recorder}
(\href{http://image-net.org/synset?wnid=n04393095}{n04393095}).
In Wikidata there is only one item.
Wikidata has multiple items for infant bed (infant bed, bassinet), the
media archive Wikimedia Commons distinguishes between baby beds,
bassinets, cradles and cribs among others.
ImageNet WordNet has two bassinet synsets, one cradle, one
\emph{carrycot} and one \emph{crib}, with four of them grouped under
\emph{baby bed}. 
ImageNet WordNet has two main \emph{mask} synsets and Wikidata has
also two items with the main label ``mask''.
They differ slightly in semantics and for the
\href{http://image-net.org/synset?wnid=n03725035}{n03725035} synset described
as ``A protective covering worn over the face'', ImageNet associates
images that are not necessarily protective coverings.
ImageNet Wordnet has \emph{cassette} and  \emph{cassette tape} synsets that
are difficult to dissociate from their description (``a container that
holds a magnetic tape used for recording or playing sound or video''
vs. ``a cassette that contains magnetic tape'').
One is grouped under \emph{container} and the other under \emph{memory
  device}.
Wikidata's cassette corresponds more to the \emph{audiocassette} of
ImageNet WordNet.

\section{Statistics}

With a SPARQL query in WDQS, we can count the number of WordNet 3.0
synsets linked from Wikidata:
\begin{lstlisting}
SELECT
  (COUNT(*) AS ?count)
WHERE {
  ?item wdt:P2888 ?uri .
  FILTER STRSTARTS(STR(?uri),
    "http://wordnet-rdf.princeton.edu/wn30/")
}
\end{lstlisting}
The response currently yields 324, i.e., approximately a third of the
total number of ILSVRC synsets.
A query for items that links to both WordNet 3.0 and BabelNet 
(with the \href{https://www.wikidata.org/wiki/Property:P2581}{P2581}
property) 
yields 105 Wikidata items. 

Wikidata items with links to WordNet synsets may have quite few
properties describing the items. 
For instance, the only properties describing \emph{oxygen mask}
(\href{https://www.wikidata.org/wiki/Q1890958}{Q1890958}) is currently
the exact match (P2888), the Wikimedia Commons category property (P373) and
3 external identifiers for Freebase, Quora and JSTOR. 
For \emph{loupe}
(\href{https://www.wikidata.org/wiki/Q4165197}{Q4165197}) there is
only one property beyond the exact match property.

We can count the number of statements (i.e., the number of property
values) for each Wikidata item with the follow WDSQ SPARQL query:
\begin{lstlisting}
SELECT
  ?item
  (COUNT(?property) AS ?count)
WHERE {
  ?item wdt:P2888 ?uri .
  FILTER STRSTARTS(STR(?uri),
    "http://wordnet-rdf.princeton.edu/wn30/")
  ?item ?property [] . 
  FILTER STRSTARTS(STR(?property),
    "http://www.wikidata.org/prop/direct/")
}
GROUP BY ?item
ORDER BY ?count
\end{lstlisting}

The sorted list shows that the WordNet 3.0-linked Wikidata items with
the most statements are typically various taxons: \emph{Dog},
\emph{bald eagle} and \emph{goldfish} each has over 100 statements. 
On the other hand, inanimate objects like  	
\emph{bathroom cabinet}, \emph{loupe}, \emph{cradle} and
\emph{cathode ray tube screen} as well as
newly created \emph{pill bottle} and \emph{bath towel} have three
or less statements, --- including the statement(s) linking to WordNet.

\begin{figure}[tb]
  \centering
  \includegraphics[width=\linewidth]{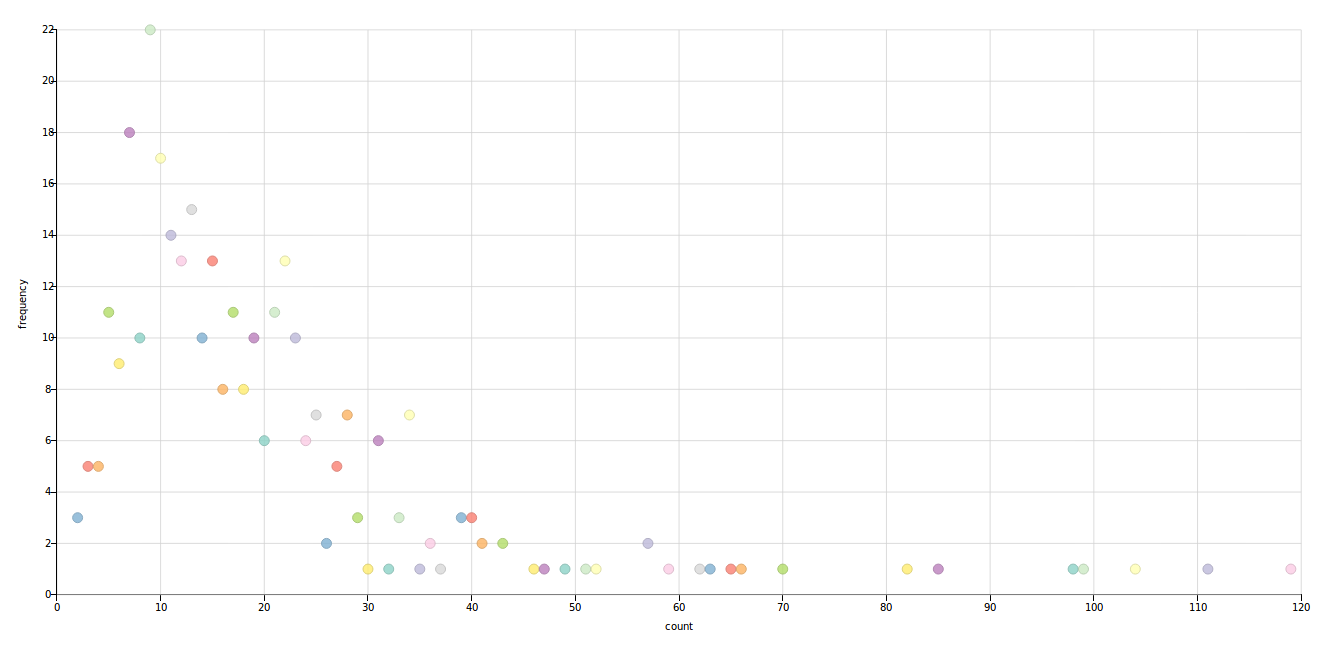}
  \caption{Plot from the Wikidata Query Service displaying the
    frequency of the number statements in each Wikidata item that is
    linked to WordNet 3.0.}
  \label{fig:statementcounts}
\end{figure}

We can get an overview of the counts by plotting the
statement counts with the following
SPARQL, producing a histogram-like plot:  
\begin{lstlisting}
SELECT 
  ?count (COUNT(?item) AS ?frequency)
WHERE {
  SELECT
    ?item
    (COUNT(?property) AS ?count)
  WHERE {
    ?item wdt:P2888 ?uri .
    FILTER STRSTARTS(STR(?uri),
      "http://wordnet-rdf.princeton.edu/wn30/")
    ?item ?property [] . 
    FILTER STRSTARTS(STR(?property),
      "http://www.wikidata.org/prop/direct/") 
  }
  GROUP BY ?item
}
GROUP BY ?count  
\end{lstlisting}
Figure~\ref{fig:statementcounts} shows the scatter plot output from
WDQS. 
The most frequent number of statements are 9.
Examples on Wikidata
items with this number of statements are \emph{filing cabinet},  	
\emph{Miniature Pinscher} (a dog bred), \emph{table lamp} and \emph{cowboy hat}

\section{Example application}

\begin{figure}[t]
  \centering
  \subfigure[``Coffee mug'']{
    \includegraphics[width=0.47\columnwidth]{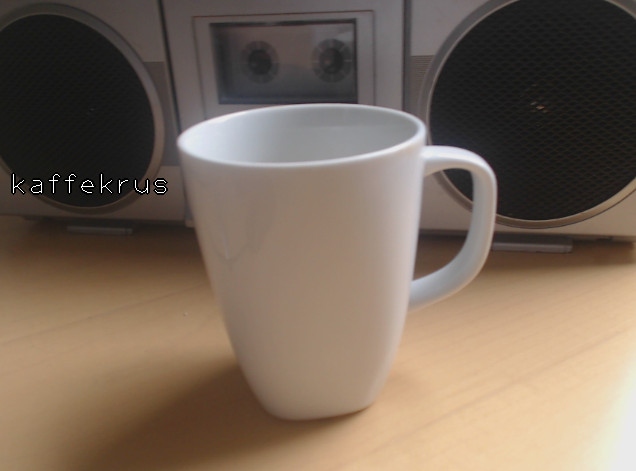}
  }
  \subfigure[``Washing machine'']{
    \includegraphics[width=0.47\columnwidth]{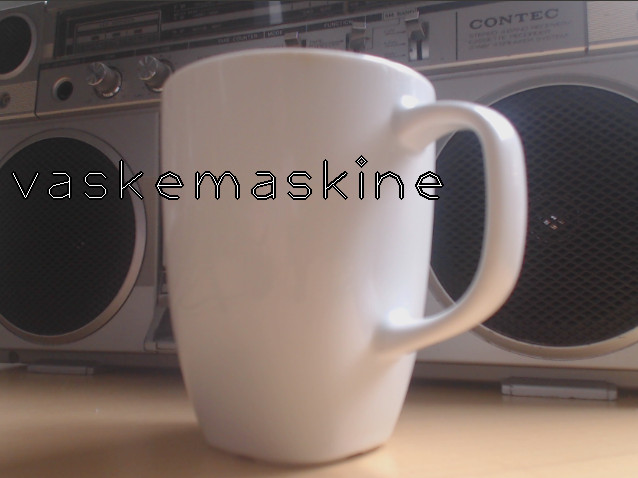}
  }
  \caption{Two screenshots of an application using the Keras Resnet-50
    model for image classification of camera images with labels
    translated to Danish via Wikidata.
    A coffee mug is in front of a tape recorder on a desk. In one case
    the label reads ``kaffekrus'' (coffee mug), in the other case---a less
    successful classification---it reads ``vaskemaskine'' (washing
    machine). 
  }
  \label{fig:application}
\end{figure}

I created an example application that would use Wikidata
in an image classification setting.
The application was implemented in Python
using the Keras deep learning framework with a TensorFlow backend.
Images were captured real-time from a USB camera with OpenCV.
For ImageNet-based image classification, I used the Resnet-50
pre-trained neural network in the form of the \verb!ResNet50! Keras
implementation.
Before the camera images were feed to the neural model, they were
cropped and preprocessed with the \verb!ResNet50! preprocessing method
following the example code provided by François Chollet at
\url{https://github.com/fchollet/deep-learning-models}. 
The index-based output of the neural network was converted to the
associated WordNet synset identifier with the decoder provided by
Keras.
Only the index associated with the highest probability in prediction
was used. 
This identifier was used for a least-recently-used cached lookup in
WDQS to identify the corresponding Wikidata item.
The Danish label from Wikidata was overlayed on the camera image and
the resulting image displayed on the computer screen via an OpenCV
function. 
The font size was dynamically adjusted to indicate the probability of
the classification.
Figure~\ref{fig:application} shows screenshots of the running application.

Running on a desktop computer with an NVIDIA Quadro K620,
the update rate with camera capture, preprocessing, neural computation
and display would take around 80 milli\-seconds for TensorFlow running on
the GPU, while around 180 milliseconds if only the CPU was used.
If the query returning the label was not cached, then the update would
take a few hundred milliseconds extra while querying WDQS.

\section{Discussion}

I have pointed to several cases of discrepancies between WordNet and
associated ImageNet images.
This is a problem if the hierarchy of WordNet or Wikidata is
used for semantic background knowledge in an image classification
setting. 
Given that ImageNet is used for benchmarking it may be less optimistic
to think that the annotation will be changed in Image\-Net.
How can Wikidata record that certain WordNet synsets are not well
represented in ImageNet?
One way would be to make a dedicated property for ImageNet synset
identifiers in Wikidata, --- independently of WordNet.
These would enable Wikidata to record ImageNet's discrepancies, e.g.,
with the ``deprecated'' feature or by a suitable qualifier attached to
the property.
A dedicated property may also make certain SPARQL query faster as it
would not require the use of the FILTER SPARQL keyword to distinguish
between different LOD URI prefixes.

The low number of statements for Wikidata items linked to ILSVRC
classes via WordNet synsets poses a problem if we want to utilize
these statements to give contexts to the categories identified with
ImageNet-trained image classifiers. 
Even for \emph{dog}---the Wikidata items with the most statements
among those items linked to WordNet currently---the number of relevant
statements to support a semantic characterization is low:
Many of the statements are for the property \emph{taxon common name}
(which is just a label)
and the number of statements which links to other statements and
characterizes the concept of \emph{dog} well may be argued to be less
than 10: It is a subclass of domesticated animal and pet; it can be
used as a pet, for hunting and guarding and as a service animal; and
it can produce a sound that is a \emph{bark}.
We may contrast that to ConceptNet where there are more statements (assertions),
but they are also less formal: According to ConceptNet a dog has a 
big heart, brains, fleas, four legs, fur, nose, one mouth, paws,
penis, teeth, two ears; a dog can be located in the backyard, in
bed, on a couch or desk, in a doghouse, etc.; and it is capable of
barking, biting and fighting a cat among lots of other
activities.\footnote{These statements 
come from conceptnet-assertions-5.5.0.csv.gz which can be downloaded
from
\url{https://s3.amazonaws.com/conceptnet/precomputed-data/2016/assertions/conceptnet-assertions-5.5.0.csv.gz},
see \url{https://github.com/commonsense/conceptnet5/wiki/Downloads}.}
There are opportunities to extend Wikidata with statements that
better characterize its items, though it is not readily clear how
this can be done.
For instance, how should we represent that a dog typically has
fur and four legs?
Or that in some contexts it forms an antonym with cat? 
One property that could be used more extensively is \emph{has quality} 
(\href{https://www.wikidata.org/wiki/Property:P1552}{P1552})
for non-material characteristics.
Physical objects may be characterized by, e.g., height and weight, but
how should we describe the size of, say, a coffee mug: By an average
height and weight over a set of coffee mugs? If yes, then where would we
get that data from? 
Wikidata items may be used to characterize other items, e.g., visual
works may be described by what they depict and written works what they
are about. 
For instance, Wikidata can describe that \emph{washing machine} is the
main topic of a set of scientific articles. 
Yet the examples we have in Wikidata now are mostly medical articles
describing accidents with washine machines. 
It not obvious how such annotations can be used with benefit in
ImageNet-based machine learning system.

Regardless of the issues around matching ImageNet synsets with
Wikidata item and the questions raised here in the discussion, I see 
interesting opportunities in the combination of machine learning and a
collaborative knowledge base such as Wikidata.


\bibliographystyle{acm}

\end{document}